\documentclass[twocolumn,superscriptaddress,aip,jcp]{revtex4}

\usepackage{amsmath,amssymb,amsfonts,bm}
\usepackage{graphicx,epstopdf}  
\usepackage{color}
\usepackage{lscape}
\usepackage{extarrows}
\usepackage{enumerate}
\usepackage{empheq}

\renewcommand{\d}{{\rm d}}

\newcommand{\w}{\omega}
\newcommand{\wti}{\widetilde}
\newcommand{\ti}{\tilde}

\newcommand{\B}{\mbox{\tiny B}}
\newcommand{\tS}{\mbox{\tiny S}}
\newcommand{\T}{\mbox{\tiny T}}
\newcommand{\Tot}{\mbox{\tiny Total}}

\newcommand{\dg}{\dagger}
\newcommand{\la}{\langle}
\newcommand{\ra}{\rangle}

\newcommand{\Sec}[1]{Sec.\,\ref{#1}}
\newcommand{\nl}{\nonumber \\}
\newcommand{\be}{\begin{equation}}
\newcommand{\ee}{\end{equation}}
\newcommand{\bsube}{\begin{subequations}}
\newcommand{\esube}{\end{subequations}}
\newcommand{\Eq}[1]{Eq.\,(\ref{#1})}
\newcommand{\Eqs}[1]{Eqs.\,(\ref{#1})}
\newcommand{\Fig}[1]{Fig.\,\ref{#1}}

\newcommand{\greater}{\mbox{\tiny $>$}}
\newcommand{\lesser}{\mbox{\tiny $<$}}

\newcommand{\sign}[1]{\mathrm{sgn}(#1)}

\begin{document}

\title{Quantum dissipation with nonlinear environment couplings: Stochastic fields dressed dissipaton equation of motion approach}

\author{Zi-Hao Chen}\thanks{Authors of equal contributions}
\affiliation{Department of Chemical Physics, University of Science and Technology of China, Hefei, Anhui 230026, China}

\author{Yao Wang} \thanks{Authors of equal contributions}
\affiliation{Hefei National Laboratory for Physical Sciences at the Microscale, University of Science and Technology of China, Hefei, Anhui 230026, China}

\author{Rui-Xue Xu}
\email{rxxu@ustc.edu.cn}

\author{YiJing Yan}\email{yanyj@ustc.edu.cn}

\affiliation{Department of Chemical Physics, University of Science and Technology of China, Hefei, Anhui 230026, China}
\affiliation{Hefei National Laboratory for Physical Sciences at the Microscale, University of Science and Technology of China, Hefei, Anhui 230026, China}
\affiliation{iChEM and Synergetic Innovation Center of Quantum Information
	and Quantum Physics,
	University of Science and Technology of China, Hefei, Anhui 230026, China}

\date{\today}

\begin{abstract}
Accurate and efficient simulation on quantum dissipation with nonlinear environment couplings
remains nowadays a challenging task.
In this work, we propose to incorporate the stochastic fields, which resolve just the nonlinear environment coupling terms,
into the dissipaton--equation--of--motion (DEOM) construction.
The stochastic fields are introduced via the Hubbard--Stratonovich transformation.
After the transformation, the resulted stochastic--fields--dressed total Hamiltonian contains only linear environment coupling terms.
On basis of that, a stochastic--fields--dressed DEOM (SFD--DEOM) can then be constructed.
The resultant SFD--DEOM, together with the ensemble average over the stochastic fields, constitutes an exact and nonperturbative approach
to quantum dissipation under nonlinear environment couplings.
It is also of relatively high efficiency and stability due to the fact that
only nonlinear environment coupling terms are dealt with stochastic fields while linear couplings are still treated as the usual DEOM.
Numerical performance and demonstrations are presented with a two-state model system.
\end{abstract}
\maketitle

\section{introduction}

Quantum dissipation is pivotal in many fields of modern science.
The underlying non-Markovian and nonperturbative quantum nature
would be prominent whenever the system and its embedded environment are highly correlated.
Various approaches have been proposed, focusing on the reduced dynamics of
system under the influence of bath.
Exact theories under Gaussian baths include the Feynman--Vernon influence functional method,\cite{Fey63118,Kle09,Wei12}
and its differential equivalence, the hierarchical--equations--of--motion (HEOM)
formalism.\cite{Tan89101,Yan04216,Ish053131,Xu05041103,Tan20020901}
Adopting dissipatons as quasi-particles to characterize the interacting bath statistical properties,
a dissipaton--equation--of--motion (DEOM) theory has been constructed.\cite{Yan14054105,Xu151816,Zha18780,Wan20041102}
The DEOM not only recovers the HEOM for the reduced system dynamics,
but also is convenient to treat the hybridized bath dynamics and polarizations.\cite{Zha15024112,Che21244105}

All these theories exploit the Gaussian thermodynamic statistics,
making them strictly valid only for the linear coupling harmonic bath.
Without loss of generality, let us consider single dissipative mode cases.
The total system--plus--bath composite Hamiltonian
takes the form,
$H_{{\T}1}=H_{\tS} + h_{\B}
+ \hat Q_{\tS}(\alpha_0+\alpha_1\hat x_{\B})$.
The system Hamiltonian $H_{\tS}$ and dissipative mode operator $\hat Q_{\tS}$
are arbitrary, whereas the bath Hamiltonian and solvation
coordinate assume
$h_{\B}= \frac{1}{2}\sum_j \omega_j(\hat p_j^2 + \hat q_j^2)$
and
$\hat x_{\B}=\sum_j c_j \hat q_j$.
In this paper, we set $\hat Q_{\tS}$ and $\hat x_{\B}$ be dimensionless.
The $\alpha$--parameters are then of energy unit.
Involved here are only $\alpha_0$--term and $\alpha_1$--term, without higher--order terms.
This linearity intrinsically implies a weak backaction
of the central system on the bath.

On the other hand, nonlinear couplings are often inevitable in real systems and
crucial in related processes.
For example, the quadratic noise fluctuations can become the
dominant source of decoherence in designing quantum computing devices.\cite{Vio02886,Mak04178301,Mul04237401,Ber05257002}
Quadratic couplings are also closely associated with the Duschinsky rotation
in studying optical spectroscopies and rate problems of molecular systems.\cite{Yan865908,Pen07114302,Wan0710369,Zha121075,Cho17074114,Wan21462}
Although there have been theoretically a few attempts to the quantum dissipative dynamics
under nonlinear bath coupling influences,\cite{Xu17395,Xu18114103,Yan19074106,Hsi18014104,Hsi20125002,Hsi20125003}
the quest of an exact quantum dissipation theory plus an efficient numerical method remains in general a
challenging task.

This paper focuses on quadratic bath coupling cases
which lead to the total Hamiltonian $H_{\T}=H_{{\T}1}+\hat Q_{\tS} \cdot \alpha_2\hat x_{\B}^{2}$
being of the form
\begin{align}\label{HT}
	H_{\T}
	=H_{\tS} + h_{\B}
	+ \hat Q_{\tS}(\alpha_0+\alpha_1\hat x_{\B}+ \alpha_2\hat x_{\B}^{2}).
\end{align}
This form of total Hamiltonian can be brought out on basis of a
widely adopted microscopic electron/exciton transfer model containing Duschinsky rotation.\cite{Xu17395,Xu18114103}
The involving bath coupling descriptors, $\{\alpha_n;n=0,1,2\}$, are found interconnected and shall be seriously determined to satisfy
basic physical requirements. This issue has been elaborated in our previous work.\cite{Xu18114103}
Meanwhile by extending the dissipaton algebra to dissipaton-pair actions,
an Ehrenfest mean-field type of DEOM approach has been constructed there for quadratic bath couplings.\cite{Xu17395,Xu18114103}

 In this work, we propose a new method to tackle the nonlinear coupling term via stochastic fields,
induced by the Hubbard--Stratonovich (HS) transformation,\cite{Str571097, Hub5977,Sha045053}
and then enrolled into the construction of DEOM.
Note that for the stochastic--fields--dressed (SFD) DEOM (SFD--DEOM) method to be developed,
the extension to include higher--order $\hat x_{\B}^{n}$--coupling terms are straightforward.
Actually in principle, the scenario can be applied to the total Hamiltonian being of
the form $H_{\Tot}=H_{\tS} + h_{\B} +\sum_a\hat Q_{a}\hat F_{a}^{n_a}$,
where $\{\hat Q_{a}\}$ and $\{\hat F_{a};\,\hat F_{a}=\sum_j c_{aj} \hat q_j\}$ are system and bath operators, respectively,
and the integers $n_a\geq 0$.
The resultant SFD--DEOM, together with the ensemble average over stochastic fields,
constitutes an exact and nonperturbative approach for quantum dissipation with nonlinear bath couplings.
The paper is arranged as follows.
Theoretical constructions are made in \Sec{thsec2},
with HS transformation in \Sec{sec2a},
SFD--DEOM construction in \Sec{sec2b},
and a norm conserved propagation via Girsanov transformation (GT) in \Sec{sec2c}.
Numerical demonstrations are given in \Sec{sec3} and the paper is summarized in \Sec{sec4}.
Throughout the paper, we set $\hbar=1$ and $\beta=1/(k_BT)$.

\section{Theory}
\label{thsec2}
\subsection{HS transformation and SFD Hamiltonian}
\label{sec2a}

According to the total composite Hamiltonian in \Eq{HT}, 
the total propagator can be recast as
\be \label{ut}
U(t)\equiv e^{-iH_{\T}t}=\lim_{N_t\rightarrow \infty}\prod_{i=1}^{N_t}e^{-iH_{{\T}1}\Delta t}e^{-i\alpha_2\hat Q_{\tS}\hat x_{\B}^2\Delta t},
\ee
with $\Delta t=t/N_t$. Here, the nonlinear $\alpha_2$--term has been extracted out at each tiny propagating time step.
Adopting the HS transformation,\cite{Str571097, Hub5977,Sha045053} it can be expressed in the form of
\begin{align}
	e^{-i\alpha_2\hat Q_{\tS}\hat x_{\B}^2\Delta t}= \!
              \sqrt{\frac{\Delta t}{2\pi}}
       \!\int\!{\d\xi}\,
               e^{-\frac{\Delta t}{2}\xi^2}
        e^{(1-i)\xi\sqrt{\alpha_2}\hat Q_{\tS}^{\frac{1}{2}}\hat x_{\B}\Delta t}.
\end{align}
Thus, the propagator in \Eq{ut} can now be obtained as the ensemble average
 over the HS--transformation induced stochastic field, $\xi_t$, as
\be \label{uu1}
U(t)={\cal M}_{\xi_t}\big\{\wti U(t;\xi_t)\big\},
\ee
with the SFD propagator
\be \label{us}
\wti U(t;\xi_t)=\lim_{N_t\rightarrow \infty}\prod_{i=1}^{N_t}e^{-i\wti H_{\T}(\xi_t)\Delta t},
\ee
and ${\cal M}_{\xi_t}$ denoting the ensemble average over the real stochastic field, $\xi_t$.
In \Eq{us}, the SFD Hamiltonian reads
\begin{align}\label{Hxy}
	\wti H_{\T}(\xi_t)=H_{0} + h_{\B}
	+\wti Q_{\tS}(\xi_t)\hat x_{\B},
\end{align}
with $H_0=H_{\tS}+\alpha_0\hat Q_{\tS}$ and
\be \label{q_cal1}
\wti Q_{\tS}(\xi_t)=\alpha_1\hat Q_{\tS}+(1+i)\xi_t\sqrt{\alpha_2}\,\hat Q_{\tS}^{\frac{1}{2}}.
\ee
Similarly, the inverse propagator can be recast as
\be \label{uu2}
U^{\dg}(t)={\cal M}_{\xi'_t}\big\{\wti U^{\dg}(t;\xi'_t)\big\},
\ee
where
\be \label{Ubar}
{\wti U}^{\dg}(t;\xi'_t)=\lim_{N_t\rightarrow \infty}\prod_{i=1}^{N_t}e^{i \wti H^{\dg}_{\T}(\xi'_t)\Delta t},
\ee
with
\be\label{Hxyp}
{\wti H}^{\dg}_{\T}(\xi'_t)=H_{0} + h_{\B}
 +\wti Q^{\dg}_{\tS}(\xi'_t)\hat x_{\B},
\ee
and
\be \label{q_cal2}
{\wti Q}^{\dg}_{\tS}(\xi'_t)=
\alpha_1\hat Q_{\tS}+(1-i)\xi'_t\sqrt{\alpha_2}\,\hat Q_{\tS}^{\frac{1}{2}}.
\ee
Note that extensions to higher-order bath couplings can just be done
via multiple HS transformations in a recursive manner.

On basis of the above elaborations [cf.\ \Eqs{uu1} and (\ref{uu2})],
the total density operator at time $t$ can be expressed as
\begin{align}
	\rho_{\T}(t)  = U(t)\rho_{\T}(0)U^{\dg}(t)
	={\cal M}_{\xi_t,\xi'_t}\big\{\ti\rho_{\T}(t;\xi_t,\xi'_t)\big\},
\end{align}
with
\be\label{hhh}
\ti\rho_{\T}(t;\xi_t,\xi'_t)=\wti U(t;\xi_t)\rho_{\T}(0){\wti U}^{\dg}(t;\xi'_t)\equiv\ti\rho_{\T}(t),
\ee
which leads to the reduced system density operator,
$\rho_{\tS}(t)\equiv {\rm tr}_{\B}[\rho_{\T}(t)]$, the following form
\be\label{HS11}
   \rho_{\tS}(t)={\cal M}_{\xi_t,\xi'_t}\big\{\ti\rho_{\tS}^{\circ}(t;\xi_t,\xi'_t)\big\},
\ee
where
$
\ti\rho_{\tS}^{\circ}(t;\xi_t,\xi'_t)\equiv [\ti\rho_{\tS}(t)+\ti\rho^{\dg}_{\tS}(t)]/2
$
with
\be\label{tirhost}
   \ti\rho_{\tS}(t;\xi_t,\xi'_t)={\rm tr}_{\B}[\ti\rho_{\T}(t;\xi_t,\xi'_t)]\equiv\ti\rho_{\tS}(t).
\ee
For brevity in later use, we have denoted $\ti\rho_{\T}(t)$ and $\ti\rho_{\tS}(t)$
for $\ti\rho_{\T}(t;\xi_t,\xi'_t)$ and $\ti\rho_{\tS}(t;\xi_t,\xi'_t)$ in \Eq{hhh} and \Eq{tirhost}, respectively.
Involved in the SFD total Hamiltonians,
\Eqs{Hxy} and (\ref{Hxyp}),
are only linear bath couplings.
The standard DEOM construction\cite{Yan14054105,Xu151816,Zha18780}
 can thus be applied to the evolution of
$\ti\rho_{\tS}(t)$, with the total Hamiltonians being \Eqs{Hxy} and (\ref{Hxyp}) for the left and right actions, respectively.
The reduced system evolution $\rho_{\tS}(t)$ is then obtained via ensemble average over the stochastic fields.

\subsection{SFD--DEOM construction}
\label{sec2b}

We are now in the position to derive the SFD--DEOM.
Let us start from
the exponential series expansion
on the bath correlation function, which serves as the common setup
for constructing DEOM/HEOM formalisms.
This expansion is based on the fluctuation--dissipation theorem,\cite{Wei12} reading
\be\label{FDT}
\la\hat x^{\B}_{\B}(t)\hat x^{\B}_{\B}(0)\ra_{\B}
=\frac{1}{\pi}
\!\int^{\infty}_{-\infty}\!\!{\rm d}\omega\,
\frac{e^{-i\omega t}J_{\B}(\omega)}{1-e^{-\beta\omega}},
\ee
with $\hat x^{\B}_{\B}(t)\equiv e^{ih_{\B}t}\hat x_{\B}e^{-ih_{\B}t}$
and the average $\la\,\cdot\,\ra_{\B}
	\equiv {\rm tr}_{\B}[\,\cdot\,e^{-\beta h_{\B}}]/%
	{\rm tr}_{\B}(e^{-\beta h_{\B}})$
both defined in the bare--bath subspace.
The involved hybridization bath spectral density $J_{\B}(\omega)$ in \Eq{FDT}
is given by\cite{Wei12}
\begin{align}\label{JB}
	J_{\B}(\omega)=\frac{1}{2}\!\int^{\infty}_{-\infty}\!\!{\rm d}t\,
	e^{i\omega t} \la[\hat x^{\B}_{\B}(t),\hat x^{\B}_{\B}(0)]\ra_{\B}.
\end{align}
It satisfies $J_{\B}(-\w)=-J_{\B}(\w)$.
The exponential series expansion
on \Eq{FDT} can be achieved by adopting a certain sum--over--poles scheme
to expand the Fourier integrand,
followed by Cauchy's contour integration.
Together with the time--reversal relation $\la\hat x^{\B}_{\B}(0)\hat x^{\B}_{\B}(t)\ra_{\B}
	=\la\hat x^{\B}_{\B}(t)\hat x^{\B}_{\B}(0)\ra_{\B}^{\ast}$,
the expansion form of bath correlation function for $t\geq 0$ is obtained as\cite{Yan14054105,Xu151816,Zha18780}
\be\label{FBt_corr}
\begin{split}
	&\la\hat x^{\B}_{\B}(t)\hat x^{\B}_{\B}(0)\ra_{\B}
	=\sum^K_{k=1}\eta_k e^{-\gamma_k t},
	\\ &
	\la\hat x^{\B}_{\B}(0)\hat x^{\B}_{\B}(t)\ra_{\B}
	=\sum^{K}_{k=1}\eta_{\bar k}^{\ast} e^{-\gamma_k t}.
\end{split}
\ee
The second expression is due to the fact that
$\{\gamma_k\}$ must be either real or complex--conjugate paired.
The associated index $\bar k\in \{k=1,\cdots,K\}$
is defined via $\gamma_{\bar k}\equiv \gamma_k^{\ast}$.

Dissipatons, with coordinates
$\{\hat f_{k}\}$,\cite{Wan20041102} can now be introduced as statistically independent quasi--particles via
\be\label{hatFB_in_f}
\hat x_{\B}=\sum^K_{k=1}  \hat f_{k},
\ee
with $\hat f_{k}(t)\equiv e^{ih_{\B}t}\hat f_{k}e^{-ih_{\B}t}$ and
\be\label{ff_corr}
\begin{split}
	\la \hat f_{k}(t)\hat f_{k'}(0)\ra_{\B}
	=\delta_{k k'}\eta_{k} e^{-\gamma_{k}t},
\\ 
	\la \hat f_{k'}(0)\hat f_{k}(t)\ra_{\B}
	=\delta_{k k'} \eta_{\bar k}^{\ast} e^{-\gamma_{k}t}.
\end{split}
\ee
Obviously, \Eq{FBt_corr} is reproduced.
Similar to original DEOM formalism, dynamical variables in SFD--DEOM are
the SFD dissipaton--augmented--reduced density operators
(SFD--DDOs):\cite{Yan14054105,Xu151816,Zha18780}
\be \label{DDO}
\ti \rho^{(n)}_{\bf n}(t)\equiv \ti \rho^{(n)}_{n_1\cdots n_K}(t)
\equiv {\rm tr}_{\B}\big[
	\big(\hat f_{K}^{n_K}\cdots\hat f_{1}^{n_1}\big)^{\circ}\ti \rho_{\T}(t)
	\big].
\ee
Here, $n=n_1+\cdots+n_{K}$ and ${\bf n}\equiv\{n_k;k=1,\cdots,K\}$, with all $n_k\geq 0$
for bosonic dissipatons.
The product of dissipaton operators inside $(\cdots)^\circ$
is \emph{irreducible}, satisfying
$(\hat f_{k}\hat f_{j})^{\circ}
	=(\hat f_{j}\hat f_{k})^{\circ}$
for boson bathp.
Each $n$--particles SFD--DDO, $\ti \rho^{(n)}_{\bf n}$(t), is specified with
an ordered set of indexes, ${\bf n}$.
For later use, we denote also ${\bf n}^{\pm}_{k}$ which differs from ${\bf n}$ only
at the specified $\hat f_{k}$-dissipaton participation number
$n_{k}$ by $\pm 1$.
The reduced system SFD density operator is just
$\ti \rho_{\bf 0}^{(0)}(t)=\ti \rho_{0\cdots 0}^{(0)}(t)=\ti \rho_{\tS}(t)$.

In \Eq{DDO}, the $\ti\rho_{\T}(t)$, as defined  in \Eq{hhh}, satisfies
\begin{align}\label{Sch_eq}
	\dot{\ti\rho}_{\T}&(t)=-i[\wti H_{\T}(\xi_t)\ti\rho_{\T}(t)-\ti\rho_{\T}(t)
                        {\wti H}^{\dg}_{\T}(\xi'_t)]
	\nl &
	=-i[H_{0}^{\times}+h_{\B}^{\times}+\wti Q^{\greater}_{\tS}(\xi_t)\hat x_{\B}^{\greater}
         -\wti Q^{\dg\lesser}_{\tS}(\xi'_t)\hat x_{\B}^{\lesser}]\ti\rho_{\T}(t),
\end{align}
where $\hat A^{\times}\equiv \hat A^{\greater}-\hat A^{\lesser}$ and
\[
\hat A^{\greater}\ti\rho_{\T}(t)\equiv \hat A\ti\rho_{\T}(t),
\qquad
 \hat A^{\lesser}\ti\rho_{\T}(t)\equiv \ti\rho_{\T}(t)\hat A .
\]
The SFD--DEOM for the time evolution of $\ti \rho^{(n)}_{\bf n}(t)$
is obtained by applying \Eq{Sch_eq} to \Eq{DDO},
followed by the standard procedure of deriving the general
DEOM formalism.\cite{Yan14054105,Xu151816,Zha18780,Wan20041102,Zha15024112,Che21244105}
During that, key steps are
the generalized Wick's theorem,\cite{Yan14054105,Xu151816,Zha18780}
\[
	\begin{split}
		\ti\rho_{\bf n}^{(n)}(t;\hat f_k^{\greater})
    &={\rm tr}_{\B}\big[\big(\hat f_{K}^{n_K}\cdots\hat f_{1}^{n_1}\big)^{\circ}\hat f_k\ti\rho_{\T}(t)\big]
	\nl	&= \ti\rho_{{\bf n}^{+}_{k}}^{(n+1)}(t)
		+\sum_{k'}n_{k'} \la \hat f_{k'}(0^+)\hat f_{k} \ra_{\B}
		\ti\rho_{{\bf n}^{-}_{k'}}^{(n-1)}(t),
		\\
		\ti\rho_{\bf n}^{(n)}(t;\hat f_k^{\lesser})
    &={\rm tr}_{\B}\big[\big(\hat f_{K}^{n_K}\cdots\hat f_{1}^{n_1}\big)^{\circ}\ti\rho_{\T}(t)\hat f_k\big]
	\nl	&= \ti\rho_{{\bf n}^{+}_{k}}^{(n+1)}(t)
		+\sum_{k'}n_{k'} \la \hat f_{k}\hat f_{k'}(0^+) \ra_{\B}
		\ti\rho_{{\bf n}^{-}_{k'}}^{(n-1)}(t),
	\end{split}
\]
and the generalized diffusion equation,\cite{Yan14054105,Xu151816,Zha18780}
\begin{align}
	{\rm tr}_{\B}\big[\big(ih_{\B}^{\times}\hat f_k\big)\ti\rho_{\T}(t)\big]
&=
	{\rm tr}_{\B}\Big[\Big(\frac{\partial}{\partial t}\hat f_k\Big)_{\B}\ti\rho_{\T}(t)\Big]
\nl &
=
    -\gamma_k{\rm tr}_{\B}\big[\hat f_k\ti\rho_{\T}(t)\big].
\nonumber
\end{align}
The final SFD--DEOM is obtained as
\begin{align}\label{DEOM}
	\dot{\ti\rho}^{(n)}_{\bf n}=
	    & -\big(iH^{\times}_{0}+\sum_k n_k\gamma_k\big)\ti\rho^{(n)}_{\bf n}
	\nl &
	-i\sum_{k}\!\big[\wti Q^{\greater}_{\tS}(\xi_t)- \wti Q^{\dg\lesser}_{\tS}(\xi'_t)\big]\ti\rho^{(n+1)}_{{\bf n}_{k}^+}
	\nl &
	-i\sum_{k}n_{k}\big[\eta_{k}\wti Q^{\greater}_{\tS}(\xi_t)
		-\eta_{\bar k}^{\ast}\wti Q^{\dg\lesser}_{\tS}(\xi'_t)\big]
	\rho^{(n-1)}_{{\bf n}_{k}^-}.
\end{align}

\subsection{Norm conserved propagation via GT}
\label{sec2c}

In principle, we can now propagate the SFD--DEOM on sampling and obtain the
reduced system dynamics, $\rho_{\tS}(t)$, with respect to \Eq{HS11}.
However, direct implementation often easily causes instability and slow convergence.
Further modification can be made by considering the norm conserved propagation.
This can be done via the Girsanov transformation (GT).\cite{Sha045053,Oks05,Ghi9078,Gat912152}
Note that $\xi_t$ and $\xi'_t$ would be both white noises in the $\Delta t\rightarrow 0$ limit.
For white--noise--fields induced stochastic processes, the GT gives
\begin{align}\label{GT}
	\rho_{\tS}(t)={\cal M}_{\xi_t, \xi'_t}\big[\ti\rho^{\circ}_{\tS}(t;\xi_t, \xi'_t)\big]
	={\cal M}_{\ti \xi_t,\ti \xi'_t}\Bigg[\frac{\ti\rho^{\circ}_{\tS}(t;\ti \xi_t,\ti \xi'_t)}{\wti\Theta(t;\ti \xi_t,\ti \xi'_t)}\Bigg],
\end{align}
with
\begin{align}\label{theta}
	\wti\Theta(t;\ti \xi_t,\ti \xi'_t) =\exp\!\bigg\{\!\!\int_{0}^{t}\!\!{\rm d}\tau
      \Big[\frac{\lambda_{\tau}^2}{2}\!-\!\lambda_{\tau}\ti \xi_{\tau}
       +\frac{\lambda_{\tau}^{\prime 2}}{2}\!-\!\lambda'_{\tau}\ti \xi'_{\tau}\Big]\!\bigg\},
\end{align}
and
\begin{align}\label{hope2}
	\lambda_{t}=\ti \xi_t-\xi_t\,,
  \qquad
	\lambda'_{t}=\ti \xi'_t-\xi'_t\,.
\end{align}
In the following, we denote  $\wti\Theta_t\equiv \wti\Theta(t;\ti \xi_t,\ti \xi'_t)$ for convenience
and choose
\be\label{tithetatcho}
\wti\Theta_t={\rm Re}\,{\rm tr}_{\tS}[\ti\rho_{\tS}(t;\ti \xi_t,\ti \xi'_t)],
\ee
for the norm conservation condition.

The problem now is to determine ($\ti\xi_t,\ti\xi'_t$) from ($\xi_t,\xi'_t$).
The stochastic fields entering \Eq{DEOM} in computation
are then ($\ti\xi_t,\ti\xi'_t$) instead of ($\xi_t,\xi'_t$).
The reduced system density $\rho_{\tS}(t)$ is then obtained via the second identity of \Eq{GT}
where $\ti \rho_{\tS}^{\circ}(t;\ti \xi_t,\ti \xi'_t)=[\ti \rho_{\bf 0}^{(0)}(t;\ti \xi_t,\ti \xi'_t)+{\rm h.c.}]/2$.
Firstly, for a single trajectory, we have, from \Eq{DEOM}, for $\wti\Theta_t$ of \Eq{tithetatcho},
\begin{align} \label{theta2}
	\dot{\wti\Theta}_t /\wti\Theta_t & ={\rm Im}\,\left\{
              \sum_k {\rm tr}_{\tS}\Big\{\big[\ti Q_{\tS}(\ti\xi_t)-\ti Q^{\dg}_{\tS}(\ti\xi'_t)\big]\ti\rho^{(1)}_{k}(t)\Big\}
                \right\}/\wti\Theta_t
	\nl                            &
	\equiv  \wti w^-_{t}\ti \xi_t+\wti w^+_{t}\ti \xi'_t\,,
\end{align}
where
\begin{align} \label{modified_noise_b}
  \wti w^\pm_{t}={\rm Re}\,\left\{(1\pm i)\sqrt{\alpha_2}\,
     \Big\{\sum_{k} {\rm tr}_{\tS}\big[\hat Q_{\tS}^{\frac{1}{2}}\ti\rho^{(1)}_{k}(t)\big]\Big\}
        \right\}/\wti\Theta_t,
\end{align}
with
\be\label{rho1k}
  \ti\rho^{(1)}_{k}(t)\equiv {\rm tr}_{\B}\big[
	\hat f_{k}   \ti \rho_{\T}(t;\ti \xi_t,\ti \xi'_t)
	\big].
\ee
Next, from \Eq{theta}, we have
\begin{align}\label{theta_dot}
	\dot{\wti\Theta}_t /\wti\Theta_t =
   \frac{\lambda_{t}^2}{2}-\lambda_{t}\ti \xi_{t}+\frac{\lambda_{t}^{\prime 2}}{2}-\lambda'_{t}\ti \xi'_{t}.
\end{align}
 Comparing \Eq{theta2} with \Eq{theta_dot}, we may set
\be
   \wti w^-_{t}\ti \xi_t=\frac{\lambda_{t}^2}{2}-\lambda_{t}\ti \xi_{t}\,.
\ee
Substituting \Eq{hope2} into the above equation gives
\be\label{uuu3}
\ti \xi_t=\sign{\xi_t} \sqrt{\xi_t^2+(\wti w^-_{t})^2}\,-\wti w^-_{t}.
\ee
Here $\sign{\cdot}$ is the sign function.
The result of $\ti \xi'_t$ can be obtained similarly.
The transformation of stochastic fields ($\ti\xi_t,\ti\xi'_t$) from ($\xi_t,\xi'_t$)
for norm--conserved trajectory propagation is thus resolved.
In numerical implementations
 both
 the originally generated stochastic fields $\xi_t$ and $\xi'_t$
 and the GT resulted $\ti\xi_t$ and $\ti\xi'_t$
 would then all be real.

We have thus finished the whole establishment of SFD--DEOM approach.
In the norm conserved propagation,
the stochastic fields entering the SFD--DEOM, \Eq{DEOM},
would be $\ti\xi_t$ and $\ti\xi'_t$.
The work flow can be outlined as follows.
\begin{enumerate}[(1)]
\item Generate two real random numbers for $\xi_t$ and $\xi'_t$
  according to the Gaussian distribution
  centered at $0$ with the width $1/\sqrt{\Delta t}$\,;

\item Perform GT to obtain $\ti\xi_t$ and $\ti\xi'_t$;     

\item Get $\wti Q_{\tS}(\ti\xi_t)$ and $\wti Q^{\dg}_{\tS}(\ti\xi'_t)$
     via \Eqs{q_cal1} and (\ref{q_cal2}), respectively, noting that $\wti Q^{\dg}_{\tS}(\ti\xi'_t) \neq [\wti Q_{\tS}(\ti\xi_t)]^{\dg}$
     since they involve different fields;

\item Perform one time-step SFD--DEOM evolution with \Eq{DEOM};

\item Repeat Steps (1)--(4) to generate one trajectory $\ti\rho_{\tS}(t;\ti \xi_t,\ti \xi'_t)$;

\item Repeat Step (5) to generate multiple trajectories;

\item Evaluate the ensemble average until convergence via \Eq{GT}.
\end{enumerate}
The inverted expression of $\xi_t$ and $\xi'_t$
depending on $\{\ti\xi_\tau; \tau\leq t\}$ and $\{\ti\xi'_\tau; \tau\leq t\}$
can not be explicitly written due to \Eq{modified_noise_b} with \Eq{rho1k}.
Norm-conserving and non-norm-conserving (without the GT step) schemes can only be compared numerically
and will be demonstrated in \Sec{sec3}.

\section{Numerical demonstrations}
\label{sec3}

\begin{figure}[!htbp]
	\includegraphics[width=0.9\columnwidth]{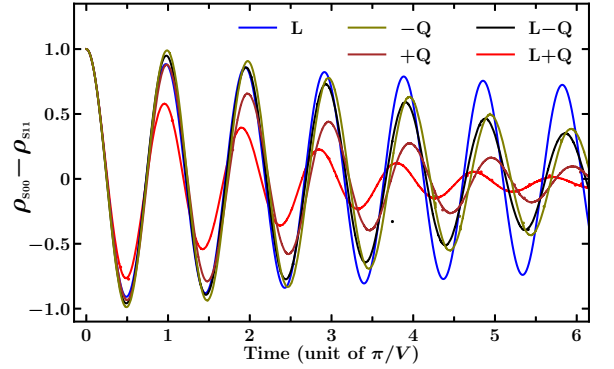}
	\caption{Population evolutions of two-state dissipative systems under
      different bath coupling cases. See the main text for the model and parameter details.}
	\label{fig1}
\end{figure}

\begin{figure}[!htbp]
	\includegraphics[width=0.9\columnwidth]{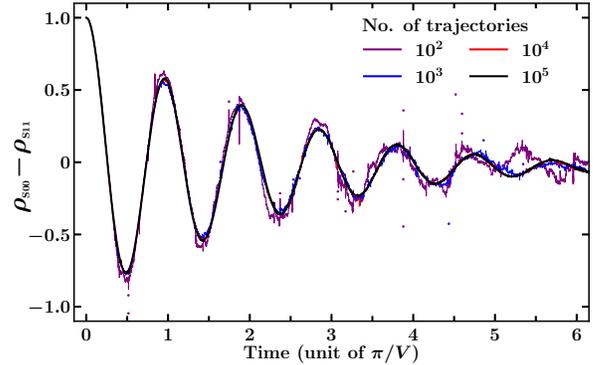}
	\caption{Time evolutions versus number of trajectories
      towards convergence of the ``L+Q''--case simulation in \Fig{fig1}.}
	\label{fig2}
\end{figure}

\begin{figure*}[!htbp]
	\includegraphics[width=0.9\textwidth]{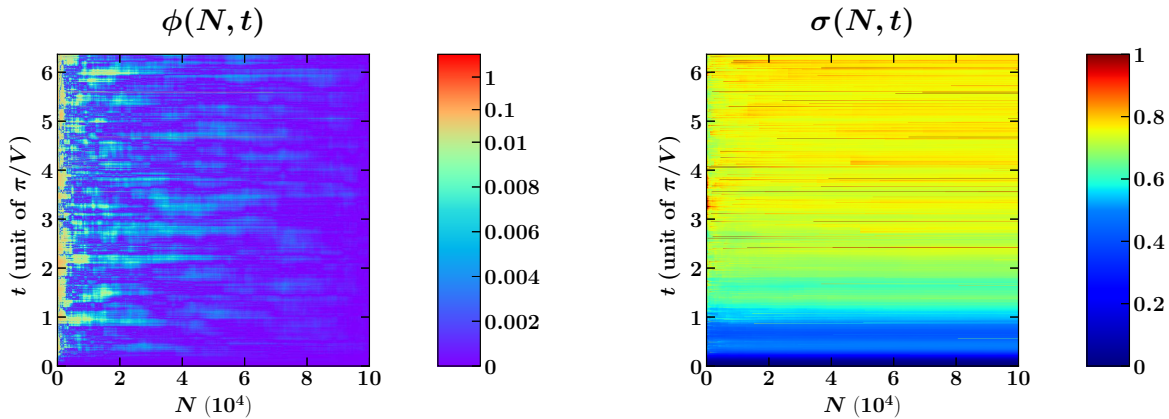}
	\caption{
     Contour plots to exhibit the convergence behaviors of average and variance versus the number
     of sampling trajectories $N$ and time $t$, $\phi(N,t)$ (left-panel) and $\sigma(N,t)$ (right-panel), respectively,
     exemplified with the ``L+Q''--case simulation in \Fig{fig1}.
     The contour coloring of the left panel uses a mixed logarithmic-rectangular scheme.
     See the main text for the definitions of $\phi(N,t)$ and $\sigma(N,t)$.
}
	\label{fig3}
\end{figure*}

\begin{figure}[!htbp]
	\includegraphics[width=0.9\columnwidth]{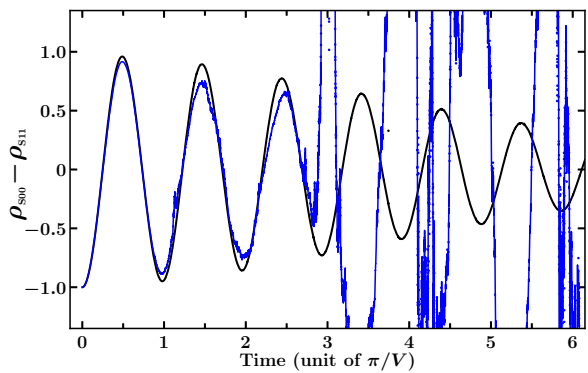}
	\caption{
     Comparison between norm-conserving (in black)
     and non-norm-conserving (in blue) calculations of \Fig{fig1}'s ``L$-$Q''--case
     upon 4$\times10^4$ trajectories.
     The blue one diverges heavily.
}
	\label{fig4}
\end{figure}

\begin{figure*}[!htbp]
	\includegraphics[width=0.9\textwidth]{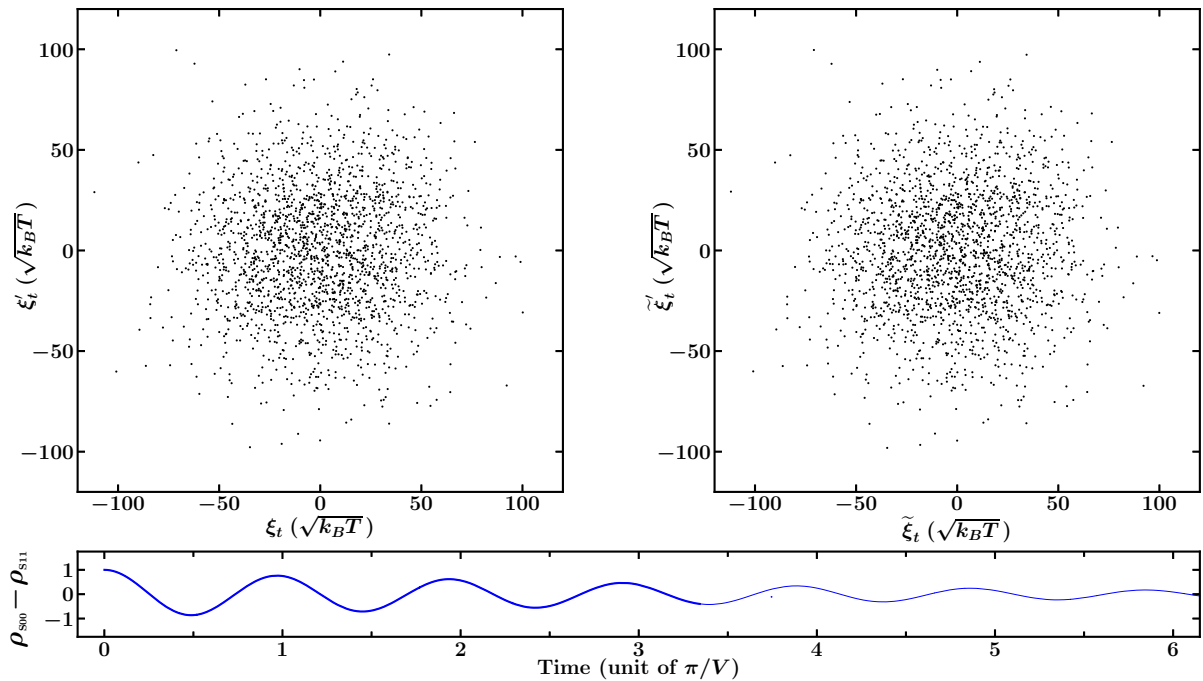}
	\caption{
     Stochastic fields before and after GT, i.e., $(\xi_t,\xi'_t)$ versus $(\ti\xi_t,\ti\xi'_t)$,
     with the time $t$ corresponding to the evolution progress bar indicated in the lower panel
     of \Fig{fig1}'s ``L$-$Q''--case.
     Points in the upper panels are drawn upon 2000 sampled trajectories from \Fig{fig4}. (Multimedia view)
}
	\label{fig5}
\end{figure*}

For numerical demonstrations, we select a two-state model system as
in Ref.\ \onlinecite{Xu18114103}. The model, corresponding to the form of \Eq{HT},
can be recast here as
\be
  H_{\tS}=\omega_{10}|1\ra\la 1|+V(|1\ra\la 0|+|0\ra\la 1|)
\ {\rm and}\
 \hat Q_{\tS}=|1\ra\la 1|.
\ee
This corresponds to the initial state being at $|0\ra$ equilibrated with the solvent before the transfer--$V$--action triggered.
Under some basic physical considerations, elaborations in Ref.\ \onlinecite{Xu18114103}
give that the $\{\alpha_0,\alpha_1,\alpha_2\}$--descriptors, which indicate the bath coupling strengths,
are related with a parameter $\theta_{\B}\equiv\omega'_{\B}/\omega_{\B}$.
Here, $\omega'_{\B}$ and $\omega_{\B}$
are the characteristic solvation--mode frequencies according to the system being at $|1\ra$ and $|0\ra$ states,
respectively.
We choose the Brownian--oscillator solvent model
\be
 J_{\B}(\omega)=\frac{\zeta\omega_{\B}\omega}{(\omega^2_{\B}-\omega^2)^2+(\zeta\omega)^2}\,.
\ee
The $\{\alpha_n\}\sim\theta_{\B}$ relations for this model are given as\cite{Xu18114103}
\be
  \alpha_0=\lambda\theta^2_{\B},
\quad
  \alpha_1=-(2\lambda\omega_{\B})^{\frac{1}{2}}\theta^2_{\B},
\quad
  \alpha_2=\frac{\omega_{\B}}{2}(\theta^2_{\B}-1).
\ee
Here, $\lambda$ is the linear--displacement induced reorganization.

In the following demonstrations, $k_BT$ is set as the unit of energy and reciprocal of time.
The other parameters are chosen as
$\omega_{10}=0$ and $V=\omega_{\B}=\zeta=1$;
$\lambda=$0.1 or 0 for with or without linear terms;
and $\theta_{\B}=0.8$, 1, 1.25 for different quadratic coupling cases.
Exhibited in \Fig{fig1} are for five conditions:
($i$) pure linear--bath--coupling (L)
      with $\lambda=0.1$ and $\theta_{\B}=1$ resulting in
      $\{\alpha_0,\,\alpha_1,\,\alpha_2\}=\{0.1,\,-0.45,\,0\}$;
($ii$) pure negative--sign ($\theta_{\B}<1$) quadratic--bath--coupling ($-$Q)
      with $\lambda=0$ and $\theta_{\B}=0.8$ resulting in
      $\{\alpha_0,\,\alpha_1,\,\alpha_2\}=\{0,\,0,\,-0.18\}$;
($iii$) pure positive--sign ($\theta_{\B}>1$) quadratic--bath--coupling ($+$Q)
      with $\lambda=0$ and $\theta_{\B}=1.25$ resulting in
      $\{\alpha_0,\,\alpha_1,\,\alpha_2\}=\{0,\,0,\,0.28\}$;
($iv$) L$-$Q
      with $\lambda=0.1$ and $\theta_{\B}=0.8$ resulting in
      $\{\alpha_0,\,\alpha_1,\,\alpha_2\}=\{0.064,\,-0.29,\,-0.18\}$;
and ($v$) L$+$Q
      with $\lambda=0.1$ and $\theta_{\B}=1.25$ resulting in
      $\{\alpha_0,\,\alpha_1,\,\alpha_2\}=\{0.16,\,-0.7,\,0.28\}$.
Apparently, for case ($i$),
SFD--DEOM is just reduced to original DEOM with no stochastic field involved.
For each of the other four cases, ($ii$)--($v$),
10$^5$ trajectories have been sampled.
Time step is set as $\Delta t$=0.001 in the unit of $(k_BT)^{-1}$.
Computing results versus number of trajectories towards convergence is illustrated in \Fig{fig2},
exemplified with the case ($v$) of ``L+Q''.
We can see that results from 10$^4$ (red) and 10$^5$ (black) trajectories almost coincide,
and that of 10$^3$ (blue) trajectories is
very close to them apart from some serration.

We may also be interested in the numerical convergence of average and variance versus the number of sampling trajectories.
Denote
$$
  P(N,t)\equiv\la{\rho_{\tS}}_{00}(t)-{\rho_{\tS}}_{11}(t)\ra_N,
$$
specifying that the average is over N trajectories.
Its variance is then defined as
$$
  \sigma(N,t)\equiv\big\la\left[{\rho_{\tS}}_{00}(t)-{\rho_{\tS}}_{11}(t)-P(N,t)\right]^2\big\ra_N^{1/2}.
$$
Introduce $\phi(N,t)\equiv\left|P(N,t)-P(N_{\rm max},t)\right|$ to show the convergence of average,
where $N_{\rm max}=10^5$ is the maximum number of trajectories in our computation.
$\phi(N,t)$ and $\sigma(N,t)$ are exhibited in the left and right panels of \Fig{fig3}, respectively.
The right panel of \Fig{fig3} demonstrates that the variance grows with $t$.
For the convergence of average, the left panel of \Fig{fig3} indicates that
more trajectories are needed for longer $t$ simulations.
The oscillating behaviors in both panels
should be caused according to the oscillation of population evolution.

The norm conservation via GT is necessary to greatly improve the sampling efficiency and simulating stability.
Non-norm-conserving calculations without adopting GT are found very hardly converged, for the cases we have tested.
The divergence of non-norm-conserving calculation is exemplified in \Fig{fig4} with the case ($iv$) of ``L$-$Q''
for the comparison between norm-conserving (in black) and non-norm-conserving (in blue) schemes,
upon $4\times10^4$ trajectories.
During the earlier period before the blue one diverges,
there is still small difference between two results.
Besides the possible reason that the black curve is converged result
while the blue one not yet,
the difference may also be caused due to
that GT is only accurate in the limit $\Delta t\rightarrow 0$
but now it is $\Delta t$=0.001 in the unit of $(k_BT)^{-1}$.
We exhibit in \Fig{fig5} (Multimedia view) the stochastic fields,
$(\xi_t,\xi'_t)$ versus $(\ti\xi_t,\ti\xi'_t)$,
drawn upon 2000 sampled trajectories from the calculations of \Fig{fig4}.
In overall speaking, the two pairs of stochastic fields, before and after GT,
are seen to be of similar distribution
with the distribution width about $1/\sqrt{\Delta t}\approx30\sqrt{k_BT}$.
Thus the GT actually does not alter the basic statistical properties of stochastic fields,
but the involved norm conservation treatment
constitutes the crucial step to successfully carry out the SFD--DEOM simulations.

\section{Summary}
\label{sec4}
In summary, we propose a stochastic--fields--dressed dissipaton--equation--of--motion (SFD--DEOM)
method to tackle the nonlinear coupling bath effects.
The stochastic fields are introduced via the Hubbard--Stratonovich (HS) transformation
just for the nonlinear bath coupling components.
After the HS transformation, the total Hamiltonian is converted to the common linear bath coupling form
and DEOM can then be constructed under the stochastic dressing fields.
Originally, dissipatons are quasi-particles characterizing the statistical
effects of linear coupling Gaussian bath.
The stochastic fields promote them to treat further nonlinear bath couplings. With the ensemble average over these fields,
the SFD--DEOM provides an exact and nonperturbative approach to quantum dissipation under nonlinear bath couplings.
Althought the paper is exemplified just with quadratic bath couplings,
the SFD--DEOM method can be systematically generalized to higher--order bath couplings via multiple HS transformations.
It can also serve as a basis for further development
of other practical simulation methods toward realistic molecular systems in condensed phases.

\begin{acknowledgements}
Support from
the Ministry of Science and Technology of China, Grant No.\ 2017YFA0204904,
and the National Natural Science Foundation of China, Nos.\ 21633006, 22103073, and 22173088
is gratefully acknowledged.
Wang Y and Chen ZH thank also the partial support from GHfund B (20210702).
\end{acknowledgements}

\vspace{2em}

\noindent{\bf Data Availability}:
The data that support the findings of this study are available from the corresponding author upon reasonable request.


\end{document}